\documentstyle[11pt]{article}
\newcommand{\be}{\begin{equation}}
\newcommand{\ee}{\end{equation}}
\newcommand{\bea}{\begin{eqnarray}}
\newcommand{\eea}{\end{eqnarray}}
\newcommand{\lb}{\label}

\begin{document}
\begin{titlepage}
\begin{flushright}
 ZU-TH 3/93\\
gr-qc/9306016
\end{flushright}
\begin{center}
\vfill
{\large\bf  Topology, Decoherence, and Semiclassical Gravity$^*$}
\vfill
{\bf Claus Kiefer}
\vskip 0.5cm
Institute for Theoretical Physics, University of Z\"urich,
Sch\"onberggasse 9,\\CH-8001 Z\"urich, Switzerland
\end{center}
\vfill
\begin{center}
{\bf Abstract}
\end{center}
\begin{quote}
We address the issue of recovering the time-dependent Schr\"{o}dinger equation
from quantum gravity in a natural way. To reach this aim it is necessary to
understand the nonoccurrence of certain superpositions in quantum gravity. We
explore various possible explanations and their relation. These are the
delocalisation of interference terms through interaction with irrelevant
degrees of freedom (decoherence), gravitational anomalies, and the possibility
of $\theta$ states. The discussion is carried out in both the geometrodynamical
and connection representation of canonical quantum gravity.
 \end{quote}
\vfill
 $^*$ Supported by the Swiss National Science
Foundation.

\end{titlepage}

 Complex numbers play a fundamental role in quantum theory.  This was not
easily accepted in the early days of quantum mechanics, and only in the last of
his seminal papers on {\em quantization as an eigenvalue problem} \cite{Sch}
did Schr\"{o}dinger introduce the $i$ in his equation.  Again, Ehrenfest
\cite{Eh} expressed his uneasiness about complex wave functions.   In his
response,
 Pauli \cite{Pa} argued  that they are unavoidable if one is interested in
conserved positive probabilities which are bilinear in the wave function and
which do not contain  its time derivatives.

A second, powerful, argument is due to Stueckelberg \cite{St} who showed that
if one starts with a {\em real} Hilbert space the uncertainty relations cannot
hold unless a new operator, $\hat{J}$, with $\hat{J}^{2}=-1$, is introduced.
The use of Hilbert spaces as a convenient mathematical framework for quantum
theory has been motivated by Wootters \cite{Wo} who argued that an appropiately
defined notion of "statistical distance", as suggested by the probabilistic
character of quantum theory, coincides with the angle between rays in an
appropriate Hilbert space.

Thus, as far as ordinary quantum theory is concerned, Ehrenfest's question was
answered in a satisfactory way. The problem, however, shows up again within the
framework of quantum gravity since there is {\em no} term containing
$i\partial/\partial t$  at a fundamental level \cite{Ba}. Although a consistent
theory of quantum gravity does not yet exist, there has been some progress in
the context of applying canonical quantization rules to general relativity.
Basically, two different approaches can be distinguished within this scheme.
The first uses the three-metric on a three-dimensional space and its extrinsic
curvature as the basic variables (the geometrodynamical formulation), the
second uses a complex $SO(3)$ connection and the densitized triad defined on
the same three-space
(the connection or, alternatively, loop space formulation).
 The central equations in both frameworks are the quantum version of the
classical constraint equations which are implemented, in the sense of Dirac, as
conditions on physically allowed wave functionals. The most important equation
is the Hamiltonian constraint equation,

\be H\Psi=0, \lb{1} \ee

\noindent where $H$ is the total Hamiltonian including both gravitational and
non-gravi\-ta\-tio\-nal fields.
In the geometrodynamical formulation, where $\Psi$ depends on the three-metric,
(\ref{1}) is referred to as the Wheeler-DeWitt equation \cite{WDW}. In the
connection formulation it depends on the complex connection referred to above.
A major development in recent years had been the discovery of {\em exact}
formal solutions to (\ref{1}) (without matter) \cite{Ro}, \cite{F1}.
If one takes for definiteness a massive scalar field for the matter part, the
Hamiltonian reads explicitly, in the geometrodynamical formulation,
  \bea H_{g} & = & -\frac{1}{2M}G_{abcd}\frac{\delta^{2}}{\delta h_{ab}\delta
h_{cd}}-2M\sqrt{h}(R-2\Lambda)
             \nonumber\\
             & &
-\frac{1}{2\sqrt{h}}\frac{\delta^{2}}{\delta\phi^{2}}+\frac{\sqrt{h}}{2}(h^{ab}\partial_{a}\phi\partial_{b}\phi
+m^{2}\phi^{2}), \lb{2} \eea

\noindent where $M=(32\pi G)^{-1}$, and $G$, $R$, $\Lambda$ are respectively
the gravitational constant, the three-dimensional Ricci scalar, and the
cosmological constant.  The coefficients $G_{abcd}$ depend on the metric and
play themselves the role of a metric on Riem$(\Sigma)$, the space of all
three-metrics. Its distinguished property is its hyperbolicity at each space
point which gives rise to an intrinsic ("many-fingered") timelike variable. We
do not consider any factor ordering terms and regularization prescriptions
since we are not interested in finding exact solutions to (\ref{1}) but in
making contact with established physics from a conceptual point of view. Since
we will remain in the realm of the semiclassical approximation, such
prescriptions can be consistently neglected.

In the connection representation, the Hamiltonian is given by the expression
\cite{F2}
  \bea H_{c} &=& \frac{G}{4\pi}\epsilon^{ijk}F_{ab}^{k}\frac{\delta^{2}}{\delta
A^{i}_{a}\delta A^{j}_{b}}
    +\frac{\delta^{2}}{\delta\phi^{2}} +
2G^{2}\partial_{a}\phi\partial_{b}\phi\frac{\delta^{2}}
      {\delta A_{a}^{i}\delta A^{i}_{b}} \nonumber\\
    & & \
+\frac{\sqrt{2}}{3}(G^{3}m^{2}\phi^{2}+G^{2}\Lambda)\eta_{abc}\epsilon^{ijk}
         \frac{\delta^{3}}{\delta A_{a}^{i}\delta A_{b}^{j}\delta A_{c}^{k}},
\lb{3} \eea
where$F_{ab}^{k}=\partial_{a}A_{b}^{k}-\partial_{b}A^{k}_{a}-\epsilon^{klm}A_{a}^{l}A_{b}^{m}$ is the field strength tensor associated with  the complex connection
${\bf A}$, and $\eta_{abc}$ is the metric-independent totally skew-symmetric
density of weight $-1$.  $F_{ab}^k$ plays the role of a metric in the space of
all connections. In contrast to (\ref{2}), it is complex and not hyperbolic.
Moreover, it is not even ultralocal.
 The important difference between (\ref{2}) and (\ref{3}) is that $H_{c}$ does
not contain the Ricci scalar, that it contains only terms with functional
derivatives (even third order derivatives) and that it is intrinsically complex
since ${\bf A}$ is complex. Note that a factor of $G$ is associated with each
functional derivative with respect to $A^{i}_{a}$ which comes with a term
containing $\phi$.

As mentioned above, the important feature about both formulations is the {\em
absence} of any external time parameter in (\ref{1}). A necessary condition for
the whole formalism is of course that it be possible to recover, at least
approximately, the familiar, and well-tested, Schr\"{o}dinger equation with its
$i\partial/\partial t$ term,  independent of whether a sensible concept of time
can be introduced at the fundamental level of Eq. (\ref{1}) itself or not. A
straightforward possibility to get the desired term would be to include a
matter field with {\em linear} momentum into the fundamental Hamiltonian. This
has been discussed through the introduction of a dust field \cite{KT}. Since,
however, this approach may create its own problems \cite{Is}, we will not take
into account such dust fields and work directly with the Hamiltonian (\ref{2})
or (\ref{3}).

  The Schr\"odinger equation can then be recovered, at least in a formal way,
through a Born-Oppenheimer kind of expansion with respect to the Planck mass
\cite{S}. This approximation scheme should make it transparent how complex
numbers enter into ordinary quantum theory.
 Let us briefly review how it works.

One starts by writing the total state in the form $\Psi=e^{iS}$ (with a complex
function $S$) and expanding $S$ in powers of the Planck mass (we use
$M\equiv(32\pi G)^{-1}$ in the geometrodynamical and $M\equiv G^{-1}$ in the
connection formulation), $S=MS_{0}+S_{1}+M^{-1}S_{2}+\ ...$. This ansatz is
inserted into Eq. (\ref{1}) and leads to  a series of equations at consecutive
orders in $M$.

We discuss first the geometrodynamical approach. The highest order ($M^{2}$)
leads to the condition that $S_{0}$  must not depend on the matter fields. If
there were, say, $N$ scalar fields $\phi_{i}$ in the matter Hamiltonian, this
equation would read

\be \sum_{i,j}\left(\frac{\delta
S_{0}}{\delta\phi_{i}}\right)\left(\frac{\delta S_{0}}{\delta\phi_{j}}\right)
g_{ij}(\phi)=0, \lb{4} \ee
where $g_{ij}$ denotes a positive definite quadratic form. It is crucial for
the success of the present approximation scheme that one can conclude from
(\ref{4}) that $S_{0}$ does {\em not} depend on the fields $\phi_{i}$. Since
non-gravitational fields possess positive kinetic energy, this conclusion can
be drawn if one requires that $S_{0}$ is either real or pure imaginary. It is
an interesting observation that, due to the indefinite nature of the
gravitational kinetic energy, one cannot derive in this scheme the opposite
limit of classical matter fields coupled to quantum gravity. This is in
contrast to, say, electromagnetism coupled  to matter where each part can be
treated semiclassically in an appropriate limit \cite{QED}.

The next order ($M^{1}$) yields the Hamilton-Jacobi equation for gravity
alone\cite{F3}
 \be \frac{1}{2}G_{abcd}\frac{\delta S_{0}}{\delta h_{ab}}\frac{\delta
       S_{0}} {\delta h_{cd}} -2\sqrt{h}(R-2\Lambda)=0. \label{5} \ee
Eq. (\ref{5}) is equivalent to Einstein's field equations \cite{Ge} and thus
corresponds to the description of a (semi)classical gravitational background.

The most important step happens in the next order ($M^{0}$) where the
time-dependent Schr\"{o}dinger equation can be recovered {\em provided} that
$S_{0}$ is chosen to be real (strictly speaking, with a negligible imaginary
part). This corresponds to the {\em choice of a complex wave function} at this
order. It happens at this point that complex numbers come into play, and it is
exactly this step that has been criticized as being ad hoc \cite{Ba}. The
procedure continues by introducing a wave functional $\chi\equiv De^{iS_{1}}$
(and demanding a certain condition on the real prefactor $D$), which obeys
\cite{F4}
 \be iG_{abcd}\frac{\delta S_{0}}{\delta h_{ab}}\frac{\delta \chi}
{\delta h_{cd}}\equiv i\frac{\delta \chi}{\delta\tau} =H_{m}\chi,
\label{6} \ee
where $H_{m}$ is the matter part of the Hamiltonian density (\ref{2}). Eq.
(\ref{6}) is the functional Schr\"{o}dinger equation or Tomonaga-Schwinger
equation for matter fields propagating in the classical background given by
(\ref{5}); $\tau({\bf x};h_{ab}]$ is a "many-fingered" time parameter. The
emergence of a time parameter in the semiclassical limit reflects the fact that
time is not an absolute entity but is inextricably entangled with the actual
world: It is defined, in this approach, by "evolving" three-geometries. It is
interesting to observe that this concept corresponds exactly to the way time is
most accurately "measured" in practice
-- to {\em ephemeris time} \cite {Ba1}. Briefly speaking, ephemeris time is
determined by inverting solutions of the equations of motion to give time as a
function of the (observed) position of a celestial body.
 All astronomical data that have been collected in the last centuries are
implemented in the definition of ephemeris time which can only be determined,
for a certain event, in retrospect.
  It is impressive to recognize that, as can be seen from the timing of the
binary pulsar PSR 1913+16 \cite{DT}, already the motion of the whole Galaxy has
to be taken into account. Eventually all available data of the cosmological
evolution will have to be implemented in the determination of ephemeris ("WKB")
time.

We will not pursue the present approximation scheme further which in the next
order leads to small correction terms to the Schr\"{o}dinger equation induced
by quantum gravity \cite{KS} showing that unitary time evolution is only an
approximate concept. We also note that there exist proposals to generalize the
above notion of semiclassical time to a full definition of time in quantum
gravity \cite{PGS}. Time is there defined {\em exactly} by the total phase of
the wave functional, but, again, this demands the presence of a complex state
as a solution to Eq. (\ref{1}).

The choice of an imaginary solution for $S_{0}$ in (\ref{5}) would lead to a
diffusion type of equation instead of the Schr\"{o}dinger equation (\ref{6}).
In quantum gravity this choice is sometimes interpreted as describing a
euclidean spacetime \cite{Ha1}. The problem discussed here is, however, not so
much related to the possible existence of euclidean worlds than to the fact
that {\em superpositions} of WKB states would not allow the recovery of the
Schr\"{o}dinger equation as presented above. The superposition
$\frac{1}{D}e^{iS_{0}}+\frac{1}{D}e^{-iS_{0}}$ (which is a  real solution to
the Wheeler-DeWitt equation (\ref{1}) at order $M$), for example, does {\em
not} lead to Eq. (\ref{6}). The above derivation has therefore been criticized
by several authors \cite{Ba,Is,Ku} since it heavily relies on the choice of a
very special, complex, state in this order of approximation.

 We now repeat briefly the above derivation of the Schr\"{o}dinger equation in
the connection representation.
Since the connection is complex, a complex structure enters the scene already
at the fundamental level of the configuration space. It is important to note
that the wave functional is a holomorphic functional of the connection
(analogously to the Bargmann representation for the harmonic oscillator). There
is, however, still the problem that only a special WKB-state allows one to
derive the Schr\"odinger equation for matter fields.
Expanding, again, $S[A^{i}_a,\phi]$ in powers of the gravitational constant and
inserting the state into (\ref{1}) with the Hamiltonian given by (\ref{3}), one
finds that $S_{0}$ does not depend on matter fields and that it obeys the
Hamilton-Jacobi equation
\be \frac{\epsilon^{ijk}}{4\pi}F_{ab}^{k}\frac{\delta S_{0}}{\delta
A_{a}^{i}}\frac{\delta S_{0}}{\delta
A_{b}^{j}}+\frac{i\sqrt{2}}{3}\Lambda\eta_{abc}\epsilon_{ijk}\frac{\delta
S_{0}}{\delta A_{a}^{i}}\frac{\delta S_{0}}{\delta A_{b}^{j}}\frac{\delta
S_{0}}{\delta A_{c}^{k}}=0. \lb{7} \ee
 Note that since the momentum conjugate to $A_{a}^{i}$, $\tilde{E_{i}^{a}}$, is
replaced by $\delta/\delta A_{a}^{i}$ (without an $i$) in the Schr\"{o}dinger
representation, the momentum is given by $\tilde{E_{i}^{a}}=i\delta
S_{0}/\delta A_{a}^{i}$. The triad $\tilde{E_{i}^{a}}$ is not necessarily real,
but it can be chosen to be real by making use of one of the remaining
constraint equations.
  Note that $S_0=constant$ is always a solution of this equation.

The next order ($G^{0}$) yields the functional Schr\"{o}dinger equation for the
wave functional $\chi\equiv De^{iS_{1}}$,
\be i\frac{\epsilon^{ijk}}{4\pi}F_{ab}^{k}\frac{\delta S_{0}}{\delta
A_{a}^{i}}\frac{\delta \chi}{\delta A_{b}^{j}}\equiv
i\frac{\delta\chi}{\delta\tau}=\tilde{H_{m}}\chi, \lb{8} \ee
where the Hamiltonian density $\tilde{H_{m}}$ is now given by the expression
\bea \tilde{H_{m}} &=&
-\frac{1}{2}\frac{\delta^{2}\chi}{\delta\phi^{2}}+\frac{\delta S_{0}}{\delta
A_{a}^{i}}\frac{\delta S_{0}}{\delta A_{b}^{i}}\partial_{a}\phi\partial_{b}\phi
 + \frac{i}{3\sqrt{2}}m^2\phi^2 \eta_{abc}\epsilon_{ijk}\frac{\delta
S_{0}}{\delta A_{a}^{i}}\frac{\delta S_{0}}{\delta A_{b}^{j}}\frac{\delta
S_{0}}{\delta A_{c}^{k}}\nonumber\\
 & & +\sqrt{2}\Lambda\eta_{abc}\epsilon_{ijk}\frac{\delta S_{0}}{\delta
A_{a}^{i}}\left(\frac{\delta S_{0}}{\delta A_{b}^{j}}\frac{\delta\chi}{\delta
A_c^k}-\frac{1}{D}\frac{\delta S_{0}}{\delta A_{b}^{j}}\frac{\delta D}{\delta
A_c^k}\chi+\frac{\delta^2S_{0}}{\delta A_b^j\delta A_c^k}\chi\right).\lb{9}
\eea
At this stage, some comments are in order.

Firstly, the Hamiltonian density  $\tilde{H_{m}}$ is equal to $H_{c}$ (\ref{3})
as evaluated on the classical gravitational background determined by the
Hamilton-Jacobi equation {\em except} for the last three $\Lambda$- dependent
terms in (\ref{9}) which arise due to the presence of the third functional
derivatives in (\ref{3}). This does not happen in the geometrodynamical
formulation.  The last two terms can be absorbed in an $\bf{A}$-dependent phase
but the first term may contain the matter field and may thus lead to a
modification of the Hamiltonian (\ref{3}). The interpretation of this term is
not yet understood.

The second point has to do with the complex nature of the connection. Eq.
(\ref{8}) can be written  as a functional Schr\"odinger equation if, in an
appropriate region of configuration space, a time functional
$\tau(\bf{x};\bf{A}]$ can be introduced such that
\be \delta({\bf x}-{\bf y})=\frac{\epsilon^{ijk}}{4\pi}F^k_{ab}({\bf
y})\frac{\delta S_{0}}{\delta A^i_a({\bf y})}\frac{\delta\tau({\bf x};{\bf
A}]}{\delta A^j_b({\bf y})}. \lb{10} \ee
This time functional, however, is in general complex. Since the wave functional
is assumed to be a holomorphic functional  in any order of approximation, its
derivative with respect to $\tau$ is fully determined by its derivative with
respect to the real part of $\tau$ which may thus serve to play the role of
physical time.
 The semiclassical expansion in this approach has, of course, also to include
an expansion of the {\em reality conditions} \cite{As} which, in highest order,
leads to a
restriction of the original configuration space onto a subspace.

Thirdly, the above derivation of the Schr\"odinger equation has to be
contrasted with the corresponding derivation in \cite{As}. A Schr\"odinger
equation  has there been found by expanding the classical Hamiltonain
constraint to second order around a chosen classical background  and then
applying this truncated constraint as a condition on the wave functional. Using
arguments similar to above, the holomorphicity of the wave functional enabled
one to identify a certain functional of the imaginary part of the trace part of
${\bf A}$ with physical time. The gravitational part {\em alone} then satisfies
a Schr\"odinger equation with respect to this time variable. This is the main
difference to our approach which only attempts to derive such an equation for
quantum matter fields in a {\em semi}classical gravitational background. The
derivation in \cite{As} is analogous to a similar derivation in the
geometrodynamical approach \cite{Ku2} where time has been constructed from the
extrinsic curvature. The similarity

is not surprising since the imaginary part of the connection  is given by the
extrinsic curvature.

How, then, can one justify the use of a single state $D^{-1}e^{iS_0}$ in the
derivation of the Schr\"odinger equation? One attempt at a possible solution
makes use of the notion of {\em decoherence} \cite{De1,De2}. The key ingredient
is the fact that only a tiny fraction of the configuration space is accessible
to observation. The unobservable degrees of freedom can thus be considered as
being irrelevant and have to be traced out. Since there exist in general {\em
quantum correlations} between all degrees of freedom, this leads to a
nonunitary master equation for the relevant system describing the suppression
of interference terms. In the geometrodynamic approach, the simplest
superposition  is (to order $M^0$)

\be \Psi\approx \frac{1}{D}e^{iS_0}\chi+\frac{1}{D}e^{-iS_0}\chi^*, \lb{11} \ee
where both $\chi$ and $\chi^*$ may be consistently assumed to satisfy the
time-dependent Schr\"odinger equation ($\chi^*$ with the sign-reversed
time-parameter).  While $S_0$ depends only on the three-metric (say, the scale
factor of the Universe in a simple minisuperspace model), $\chi$ depends on the
three-metric as well as on all nongravitational fields. The idea is that most
of these degrees of freedom contained in $\chi$ are inaccessible. The relevant
object on which one has to focus is thus the reduced density matrix for the
gravitational part,

\be \rho[h_{ab},h_{ab}']={\mbox
Tr}_{\phi}\Psi^*[h_{ab}',\phi]\Psi[h_{ab},\phi],  \lb{12} \ee
where $\phi$ symbolically denotes the fields which are traced out.
Decoherence then occurs if in the total density matrix
\[
\Psi^*\Psi=\frac{1}{D^2}(2\vert\chi\vert^2+e^{2iS_0}\chi^2+e^{-2iS_0}\chi^{*2})
\]
the last two terms become small after integration over the $\phi$-variables.
The reduced density matrix then describes an approximate {\em ensemble} of the
two states in (\ref{11}).
In a typical model one takes for $\phi$ the inhomogeneous modes of a scalar
field and  restricts attention only to the radius of the Universe as the
relevant degree of freedom. For the superposition (\ref{11}) one can  then find
a suppression of interferences.  It has been found, for example, that the
suppression factor is proportional to $\exp(-\pi mH_0^2a^3/128)$, where $m$,
$H_0$, $a$ are respectively the mass of the scalar field, the Hubble parameter,
and the scale factor \cite{De2}. Except for small scales and near the turning
point of the corresponding classical Universe one can consistently treat the
two components in (\ref{11}) as being dynamically independent. One has to
emphasize, of course, that the degree of decoherence depends on the total state
and on the choice of relevant and irrelevant variables.  As the discussed
examples indicate, decoherence should become effective if a huge number of
degrees of freedom as well as large masses (in our case it is the large Planck
mass) are involved.

In the general case one has, instead of (\ref{11}), many WKB components,
\[ \Psi\approx\sum_r e^{iS_r}\chi_r, \]
where each component has its own WKB time. Decoherence then should explain why
the second sum in the total density matrix,
\[ \Psi^*\Psi\approx \sum_r\vert\chi_r\vert^2+\sum_{r\neq
s}e^{i(S_r-S_s)}\chi_s\chi^*_r , \]
becomes negligible after integrating out the irrelevant degrees of freedom.

Independent of this approach through decoherence it has been suggested to
introduce complex numbers on the  level of (\ref{1}) itself so that it will be
natural to start with a complex WKB state from the very beginning. This can be
achieved, for example, if the functional derivative in (\ref{2}) is replaced
according to the prescription \cite{Ba}

\be \frac{\delta}{\delta h_{ab}} \to \frac{\delta}{\delta h_{ab}} -i {\cal
A}^{ab}[h_{cd}], \lb{13} \ee
where ${\cal A}^{ab}$ are the components of some "super- gauge potential"
defined on the configuration space (an analogous proposal can be made with
respect to (\ref{3})--see below). How could one justify the introduction of
such a super gauge potential? We provide three different answers which are,
however, not independent of each other and which may also be connected to the
notion of decoherence discussed above.

The first possibility is connected with the occurrence of {\em anomalies}. It
is well known that, for example, Weyl fermions in an external electromagnetic
background may acquire an anomaly leading to the violation of gauge invariance
(see, e.g., \cite{FJ}). This, in turn, is intimately connected with the
emergence of a "functional" Berry phase $\gamma$,

\be \gamma=\oint DA^a{\cal A}_a({\bf x};{\bf A}], \lb{14} \ee
where ${\bf A}$ denotes the usual vector potential, and

\be {\cal A}_a({\bf x};{\bf A}]=i\langle\Psi\vert\frac{\delta}{\delta A^a({\bf
x})}\vert\Psi\rangle \lb{15} \ee
is a super- gauge potential defined on the configuration space of all vector
potentials. It stands in close analogy to the potential ${\cal A}^{ab}$ in
(\ref{13}). The state $\Psi$ in (\ref{15}) is a fermionic state which also
depends on the external electromagnetic field. One can restore gauge invariance
(but loses Lorentz invariance) if one replaces the electromagnetic field
momentum according to \cite{Se}

\be \frac{\delta}{\delta A^a}\to  \frac{\delta}{\delta A^a} - i{\cal A}_a({\bf
x};{\bf A}]. \lb{16} \ee
This introduces complex numbers into the electromagnetic field Hamiltonian.

The standard model of gauge theories, however, does not possess such anomalies,
at least perturbatively. Are there gravitational anomalies of this kind? It has
been shown that the determinant of the Weyl operator for handed fermions in a
gravitational background is not invariant under frame rotations if the
dimension $d$ of spacetime is $d=4n+2=2,6,10,...$ \cite{AGW}. Such {\em Lorentz
anomalies} thus do not exist in four spacetime dimensions. An external
gravitational field influences chiral anomalies in four dimensions but not in
the standard model. Thus, as far as canonical quantum gravity in three space
dimensions is concerned, anomalies do not seem to be a realistic option to
justify the substitution (\ref{13}) (they may become relevant in the framework
of Kaluza-Klein theories).

Before we embark on a more promising explanation we note that the presence of
{\em torsion} \cite{To} could become important in this context. The action for
a Dirac field in an external gravitational field, for example, is complex if
torsion is nonvanishing. This in turn would also lead to a complex Hamiltonian
and could thus also produce a nontrivial Berry phase in analogy to the case
discussed above.

Perhaps the most promising attempt to justify a substitution like (\ref{13})
arises from the possibility to have $\theta$ {\em states} in quantum gravity
\cite{Is2}. Since the situation in the geometrodynamical formulation is
somewhat different from the connection formulation, we treat both cases
separately.  $\theta$ states can arise from the possibility to have {\em large
symmetry transformations}, i.e., symmetry transformations which are not
connected with the identity. In general relativity,  the relevant symmetry is
the invariance under the group of diffeomorphisms. The momentum constraints
secure only the invariance of the wave functional under infinitesimal
diffeomorphisms \cite{Hi} that are asymptotically trivial \cite{F5}.   One thus
has to deal with $\theta$ states if the diffeomorphism group is not connected,
i.e., if $\pi_0(\mbox {Diff}\Sigma)\equiv \mbox {Diff}\Sigma/\mbox
{Diff}_{id}\Sigma$ is non-vanishing. One would thus expect, in analogy to
ordinary quantum theory, that the wave functional

transforms according to a one-dimensional, irreducible, representation of
$\pi_0(\mbox {Diff}\Sigma)$,
and the $\theta$ sectors are then labelled by the elements of the group
Hom$(\pi_0(\mbox {Diff}\Sigma), U(1))$, i. e., the group of homomorphisms of
$\pi_0$ into $U(1)$. It has to be emphasized that the restriction to such
representations is an assumption which one might have to generalize after a
better understanding of quantum gravity will have been achieved. We note,
however, that higher dimensional representations appear even in standard
quantum mechanics in the presence of discrete groups \cite{Ni}.

The emergence of a $\theta$ parameter is well known from Yang-Mills theories
(see, e.g., \cite{Ja}). For the gauge group ${\cal G}:S^3\to SU(N)$, e. g., one
has $\pi_0({\cal G})=\pi_3(SU(N))=Z$. Thus, $\theta$ states are simply labelled
by the elements of\\ Hom$(Z,U(1))$. In the connection approach to quantum
gravity, the situation will be analogous, as discussed below. Instead of taking
the gauge group as the starting point, one can take an alternative viewpoint
and focus on the topological properties of the physical {\em configuration
space}, $Q$, of the theory. In the Yang-Mills case this is the space of vector
potentials modulo ${\cal G}$, while in gravity this is the space of Riemannian
metrics, Riem$\Sigma$, on $\Sigma$ modulo diffeomorphisms.  Quantum theory on
nontrivial configuration spaces has been extensively discussed in the
literature (see, e. g., \cite{SB}). It was found that $\theta$ structures may
emerge if the first fundamental group, $\pi_1(Q)$, is nonvanishing. The quantum
mechanical propa

gator, $K(q_2,q_1)$, of a system can then be expressed as a sum of propagators
$K_{[p]}$ where in each $K_{[p]}$ the paths lie in the same homotopy class.
Thus,

\be K(q_2,q_1)=\sum_{[p]}\chi([p])K_{[p]}(q_2,q_1), \lb{17} \ee
where $\chi\in\mbox {Hom}(\pi_1(Q),U(1))$. In the functional Schr\"odinger
equation wave functionals can be viewed as cross sections of a complex bundle
over $Q$ which gives rise to a {\em connection} over $Q$ precisely in the way
as it was envisaged in (\ref{13}). One considers flat connections only since
there is no hint of a "super connection" with non-vanishing curvature in the
framework of the Wheeler-DeWitt equation.

Both viewpoints can be unified if the symmetry group acts freely on
Riem$\Sigma$, since then $\pi_1(Q)$ is homeomorphic to $\pi_0({\cal G})$ and
the $\theta$ parameter can be directly connected with the nontrivial structure
of the configuration space itself. In gravity, however, Diff$(\Sigma)$ does not
act freely because of the existence of isometries.  These can be removed if one
goes to the "extended superspace" \cite{ES,Ni2}.
(We emphasize that the compact and open cases can be discussed on the same
footing within this framework since the corresponding configuration spaces are
 diffeomorphic.)
 $\theta$ states are then classified by the elements of Hom$(\pi_1(\mbox
{Riem}\Sigma/\mbox {Diff}_{*}\Sigma)\equiv\pi_0(\mbox {Diff}_{*}\Sigma),U(1))$,
where $\mbox {Diff}_{*}\Sigma$ is the subgroup of $\mbox{Diff}\Sigma$
consisting of diffeomorphisms which leave the tangent space at some fixed base
point invariant. It acts freely on $\mbox{Riem}\Sigma$ \cite{Is2}.
Even if the action of the group is not free, one can discuss $\theta$ sectors
which correspond to representations of the group acting on configuration space
\cite{GL}. One can even find such sectors if the first fundamental group is
vanishing \cite{Ha}.

 It is sometimes suggested that  the configuration space has to be reduced
further from three degrees of freedom per space point to two degrees of freedom
by extracting some intrinsic time variable \cite{Is2}.  This would amount to
solve the Hamiltonian constraint {\em before} quantization through the choice
of an appropriate time function. The corresponding framework would be different
from the one considered here since there would be no equation like (\ref{1}).
As far as the topological properties of the new configuration space are
concerned, one would, however, not expect drastic modifications since
$\mbox{Riem}\Sigma$ is in that case only factored by the multiplicative group
of conformal factors which is a topologically trivial operation \cite{FM}.

The most interesting question of course is: Which three-manifolds $\Sigma$ can
lead to a nontrivial $\theta$ structure? Since $\pi_0(\mbox
{Diff}_{*}S^3)=\mbox{id}$ for orienting preserving diffeomorphisms, there is no
such structure in the simplest case of a three-sphere, $\Sigma=S^3$, except for
the case that reflections are included. Nontrivial structures emerge, e. g., in
the case $\Sigma=S^1\times S^2$ ("wormhole") and $\Sigma=S^1\times S^1\times
S^1$ (the three-torus) \cite{Is2}.

The next question is: Might the presence of a $\theta$ structure be the reason
why one can focus consistently on complex solutions to (\ref{1}) and, moreover,
to only one WKB component $D^{-1}e^{iS}$ in the semiclassical approximation?
The first part of this question can be answered easily. If $\pi_0\neq0$, and if
the group is represented by a one-dimensional, irreducible, representation, the
superposition (\ref{11}) does not belong to the class of allowed quantum states
since one component would transform, under the assumption that it is an
eigenstate of a large symmetry transformation, with the complex conjugate of
the other (always assuming, of course, that the $\theta$ parameter is
nonvanishing). The state (\ref{11}) would thus not be an eigenstate of a large
symmetry transformation. This would correspond to an {\em exact} superselection
rule -- much stronger than the approximate notion of decoherence discussed
above from which it could in principle be distinguished. We note that $\theta$
is a parameter

 which can in principle be measured (in QCD, for example, it can be determined
by measuring the electric dipole moment of the neutron).

The second part of this question is more difficult to answer.  A single
semiclassical state is in general not, even within the limits of the
semiclassical approximation,  an eigenstate of a large symmetry transformation.
If the $\theta$ structure indeed gave rise to an exact superselection rule, one
would have to {\em select} amongst all possible WKB solutions those which are
eigenstates of a large symmetry transformation. This could turn out to be a
viable principle in finding physically relevant solutions since most
eigenstates are superpositions of several WKB components \cite{GL}. Because of
the complexity of the problem, concrete examples have to be discussed within
the context of very simple minisuperspace models. Consider, e. g., a model
which is defined in {\em three} spacetime dimensions, with spatial sections
$\Sigma=S^1\times S^1$ \cite{GL,LT}, by the metric

\be ds^2=-N^2(t)dt^2 + a^2(t)dx^2 + b^2(t)dy^2, \lb{18} \ee
where $x$ and $y$ are identified periodically with period $2\pi$. The
Wheeler-DeWitt equation reads, for this model,
\be H\psi\equiv\left(\frac{\partial^2}{\partial a\partial b}+\frac{\pi^2\Lambda
ab}{4G^2}\right)\psi=0,
    \lb{19} \ee
where $G$ and $\Lambda$ denote again the gravitational constant and the
cosmological constant, respectively. The minisuperspace analogue of
$\pi_0(\mbox {Diff}\Sigma)$ is here played by the permutation group of the two
scale factors, which has only two one-dimensional, irreducible,
representations. Note that the coupling to matter only produces additional
terms containing the combination $ab$ so that this symmetry will not be
spoiled. Semiclassical solutions are of the form

\be \psi=Ce^{iS}, \lb{20} \ee
where $S$ is a solution to the Hamilton-Jacobi equation
\be -\frac{\partial S}{\partial a}\frac{\partial S}{\partial
b}+\frac{\pi^2\Lambda ab}{4G^2}=0. \lb{21} \ee
In \cite{GL,LT} interest was focused on the class of solutions that correspond
to the "no-boundary proposal." These states are not eigenstates of the
permutation operator so one has to superpose two semiclassical states of the
form (\ref{20}) to arrive at states which are either symmetric or antisymmetric
under the action of the permutation group. Such superpositions, however, do
{\em not} allow one to recover the Schr\"odinger equation for matter fields.
This is possible if one chooses one of the two solutions
\be
\psi_{\pm}(a,b)=\left(\frac{1}{a}+\frac{\beta}{b}\right)\exp\left(\pm\frac{i\pi}{2G}\sqrt{\Lambda}ab\right)
    \lb{22} \ee
with either $\beta=1$ (symmetric state) or $\beta=-1$ (antisymmetric state). It
is important that the WKB prefactor can be chosen in such a way that the wave
function acquires the desired transformation properties. Note that (\ref{22})
is an exact solution of (\ref{19}). It is an artifact of this simple model,
which allows only real representation to occur, that the superposition of the
two states $\psi_+$ and $\psi_-$, which does {\em not} allow the derivation of
the Schr\"odinger equation, is again an eigenstate of the permutation operator.
In the generic case of complex representations this is no longer possible. In
the nontrivial examples in \cite{GL} only the trivial representations are
possible {\em if} single WKB states are required to be eigenstates.

The situation is somewhat different from the point of view of the connection
representation. Since the wave functional is defined on a space of connections,
one has, {\em in addition} to the diffeomorphism group, a $SO(3)$ gauge group
\cite{F7},  and hence one may have $\theta$ states in analogy to the Yang-Mills
case\cite{ABJ} independent of whether the action of the diffeomorphism group is
represented trivially or not. As discussed above, these states arise since
$\pi_0({\cal G})=Z$. Thus, as long as $\theta\neq0$, the wave functional
transforms under a {\em large} gauge transformation with winding number $n$ as

\be \Psi[{\bf A}] \to e^{in\theta}\Psi[{\bf A}]. \lb{23} \ee
Superpositions like (\ref{11}) are thus "forbidden" if one assumes, in analogy
to the Yang-Mill case, that physical states are eigenstates of the  operator
which generates large gauge transformations. This would correspond to the
decomposition of the full theory into inequivalent superselection sectors. To
say that superpositions are "forbidden" is equivalent to saying that they are
{\em indistinguishable} from an ensemble of states living in different sectors
if only gauge-invariant observables are available \cite{La}. Whether
gauge-noninvariant observables, which can bridge between these sectors, are
present at a more fundamental level, is an open question. The situation is
analogous to QED where one might speculate whether the charge superselection
rule has its prime origin in the invariance of the theory under rigid gauge
transformations or in decoherence \cite{De2,Ze}. A further example is provided
by the spontaneous symmetry breaking in the early Universe \cite{Ze} where the
various  "true vacua" are s

eparated by a superselection rule.

 Instead  of demanding that the wave functional transforms like in (\ref{23}),
one can alternatively demand that it be invariant but that instead the field
momentum changes according to \cite{ABJ}

\be \frac{\delta}{\delta A^i_a} \to  \frac{\delta}{\delta A^i_a} -
\frac{i\theta G^2}{8\pi^2}\tilde{B}^a_i,
    \lb{24} \ee
where $\tilde{B}^a_i\equiv\eta^{abc}F_{cbi}$ is the "magnetic field"
corresponding to $A^i_a$.  The gravitational constant enters explicitly in
(\ref{24}). Note that this is just the desired substitution (\ref{13}).

To conclude, we have presented some possible avenues which may lead to a deeper
understanding of the derivation of the Schr\"odinger equation from quantum
gravity.
A non vanishing $\theta$ parameter could lead to a strong selection principle
amongst possible WKB solutions. Moreover, it might be that at a fundamental
level decoherence and $\theta$ sectors are actually related, i. e., that the
$\theta$- superselection rule is {\em caused} by decoherence in a dynamical
way. A detailed discussion would, however, need to invoke a specific
"environment" which is able to decohere different $\theta$ eigenstates.
 Furthermore, since the structures of the respective configuration spaces in
the geometrodynamical and the connection dynamical formulations are different,
semiclassical considerations could perhaps provide a way to distinguish between
both approaches to quantum gravity. We hope to address these issues in a future
publication.

  It should be emphasized that the whole paper rests on the assumption that
general relativity is the starting point for quantization. Higher derivatives,
for example, would  change the situation and lead to a Schr\"odinger equation
even at the fundamental level of full quantum gravity  \cite{Ho}.

It is interesting to see that the possible connection of symmetries with the
$i$ in the Schr\"odinger equation is similar to the one envisaged in \cite{Ya},
although it emerges here at a much more fundamental level.

\vspace{2mm}

\begin{center}
{\bf Acknowledgements}
\end{center}
I owe my thanks to Julian Barbour for many discussions and for motivating me to
look more deeply into these questions. I am also grateful to Heinz-Dieter
Conradi, Domenico Giulini, Petr H\'aj\'{\i}\v{c}ek, Slava Mukhanov, J\"urgen
Tolksdorf, Andreas Wipf, and H.-Dieter Zeh for helpful discussions and critical
comments.


\begin{thebibliography}{999}
\bibitem{F3}An expansion with respect to $G$ does not allow to consider
"macroscopic" sources in (4). They can be included in the description if the
expansion if performed with respect to some other parameter ("large mass") of
the theory.
\bibitem{F4}We do not consider factor ordering terms which may arise from the
coupling of gravity to the kinetic matter terms, see Ref.~21.
\bibitem{F5} If one allows for rotations at spatial infinity, there is also a
possibility to get half-integer spin states in case that a $2\pi$ rotation acts
nontrivially, see Ref.~37 (the manifold $\Sigma=R^3\# T^3$, e. g., allows such
states). In this paper we restrict ourselves to the compact case.
\bibitem{F7} One might again be concerned about the complex nature of the
connection in this context. Since the wave functional is, however, holomorphic
with respect to the connection, it is determined by its restriction to its real
part, see Ref.~48.
\end{thebibliography}
\end{document}